%% file: main.tex
\documentclass[aip,jcp,amsmath,amssymb,superscriptaddress,reprint]{revtex4-2}
\usepackage{graphicx}
\usepackage[english]{babel}
\usepackage{xcolor}

\usepackage[
    colorlinks=true,
    citecolor=black,
    linkcolor=black,
    urlcolor=black
]{hyperref}

\begin{document}

\title{Beyond the Ground State: Assessing the Machine-Learned Skala Functional in $\Delta$SCF Calculations of Molecular Excitation Energies}

\author{Egor Trushin}
\email{e.trushin@hzdr.de}
\affiliation{Center for Advanced Systems Understanding (CASUS), G\"orlitz, Germany}
\affiliation{Helmholtz-Zentrum Dresden-Rossendorf (HZDR), Dresden, Germany}

\author{Thomas D. K\"uhne}
\affiliation{Center for Advanced Systems Understanding (CASUS), G\"orlitz, Germany}
\affiliation{Helmholtz-Zentrum Dresden-Rossendorf (HZDR), Dresden, Germany}
\affiliation{Institute of Artificial Intelligence, Technische Universit\"at Dresden, Dresden, Germany}

\date{\today}
\begin{abstract}
We assess the transferability of the machine-learned Skala exchange-correlation (XC) functional beyond ground-state energetics using molecular vertical excitation energies and singlet–triplet gaps calculated with the $\Delta$-self-consistent-field ($\Delta$SCF) approach. We compare Skala with PBE, r$^2$SCAN, and PBE0 on subsets covering small-molecule excitations (QUEST1), inverted singlet–triplet gaps (ST-gap), and intramolecular charge-transfer excitations (QUEST6). Both spin-unrestricted Kohn–Sham (UKS) and restricted open-shell Kohn–Sham (ROKS) formalisms are considered. On QUEST1, Skala achieves the lowest mean absolute errors (MAEs) for singlet and triplet excitation energies with both formalisms, although PBE0 and r$^2$SCAN provide more accurate singlet-triplet gaps through error cancellation. For ST-gap, Skala yields the lowest gap MAEs with both UKS and ROKS. Evaluating the mixed-spin and triplet energy components on the same orbitals optimized for the ROKS open-shell singlet yields negative common-orbital splittings for seven of eight ST-gap molecules with Skala. Separate optimization of the triplet orbitals increases the Skala gaps, but six remain negative. In contrast, all three conventional functionals yield positive common-orbital splittings and gaps after separate triplet optimization for all eight molecules. For QUEST6, treated with UKS only, Skala yields more accurate singlet excitation energies than PBE and r$^2$SCAN but remains less accurate than PBE0. Across QUEST1 and ST-gap, Skala yields lower MAEs with UKS than ROKS. Overall, these results support Skala's transferability to molecular excited states while highlighting its dependence on the excitation class and open-shell treatment.
\end{abstract}

\maketitle

\section{Introduction \label{sec:introduction}}

Density-functional theory (DFT) \cite{Hohenberg1964,Kohn1965,Parr1989}, within the Kohn–Sham (KS) \cite{Kohn1965} or generalized KS (GKS) \cite{Seidl1996} framework, is widely used for electronic structure calculations in chemistry, physics, and materials science owing to its favorable balance between computational cost and accuracy. The accuracy of both KS and GKS methods, however, depends on the quality of the approximation for the exchange–correlation (XC) functional. The development of more accurate XC energy functionals has therefore been a central goal of DFT research. A major strategy toward this goal has been the systematic incorporation of increasingly sophisticated ingredients into XC approximations, as illustrated by Jacob's ladder \cite{Perdew2001,Perdew2013}. The first rung corresponds to the local-density approximation (LDA), which depends only on the electron density $n(\mathbf{r})$. Generalized gradient approximations (GGAs) on the second rung additionally incorporate the density gradient $\nabla n(\mathbf{r})$, while meta-GGAs on the third rung introduce further ingredients, such as the kinetic-energy density $\tau(\mathbf{r})$, the density Laplacian $\nabla^2 n(\mathbf{r})$, or both \cite{Tao2003}. The fourth rung incorporates exact exchange, an explicitly nonlocal functional of the occupied orbitals, which can be treated using either a nonlocal Fock-type exchange operator or a local multiplicative potential obtained through the optimized effective potential method \cite{Kummel2008}. The fifth rung additionally introduces dependence on unoccupied orbitals and their energies through the correlation energy \cite{Ren2012,Goerigk2014}. Moving up Jacob's ladder generally increases computational costs, although higher-rung approximations do not necessarily provide uniformly improved accuracy across all systems and properties \cite{Mardirossian2017}.

Conventional XC functionals have often been developed using physically motivated functional forms guided by exact constraints, as exemplified by PBE \cite{Perdew1996} and SCAN \cite{Sun2015}. Machine learning (ML) offers a complementary strategy, employing highly flexible functional representations whose parameters are optimized using reference data \cite{Dick2020,Kasim2021,Bystrom2022,Gao2025}. This flexibility opens up new possibilities for XC functional design, particularly in how established electronic-structure ingredients can be combined. Skala \cite{Luise2026}, a recently developed machine-learned XC functional, illustrates one such possibility. Unlike conventional approaches that introduce nonlocal XC dependence through explicit orbital expressions, such as exact exchange, Skala learns this dependence by combining semilocal meta-GGA ingredients evaluated at different spatial points. Trained primarily on extensive and diverse high-accuracy ground-state quantum-chemical data, Skala achieves accuracy competitive with hybrid functionals for main-group energetics while retaining the favorable computational scaling of semilocal DFT.

The transferability of machine-learned XC functionals beyond their original training domains requires careful evaluation, as illustrated by SCF convergence difficulties reported for DM21 in transition-metal systems \cite{Kirkpatrick2021,Zhao2024}. Although Skala was developed primarily for ground-state molecular energetics \cite{Luise2026}, recent applications have demonstrated its transferability to periodic solids, including molecular-crystal lattice energies and electronic band gaps, without retraining \cite{Pototschnig2026}. Beyond its application to periodic solids, state-specific SCF calculations of molecular excited states \cite{Restaino2026} offer a distinct test of Skala's transferability. Since excited-state solutions were not explicitly targeted during training \cite{Luise2026}, the accuracy of excitation energies obtained from differences between ground- and excited-state total energies provides a quantitative assessment of Skala's ability to describe electronically excited states in terms of their energies relative to the ground state.

To perform this assessment, we employ state-specific orbital-optimized DFT for excited states, with excitation energies evaluated using the $\Delta$-self-consistent-field ($\Delta$SCF) approach \cite{Hait2020,Restaino2026}. In this approach, excited states are targeted starting from non-Aufbau orbital occupations \cite{Gilbert2008,Barca2018}, with the ground- and excited-state orbitals optimized in separate SCF calculations at a common nuclear geometry. For single-electron excitations from closed-shell systems, the resulting configurations contain two singly occupied orbitals that can couple to form singlet or triplet states. For open-shell singlets, a single-determinant description generally leads to spin-symmetry breaking and significant spin contamination, which can be addressed using spin purification or spin-adapted approaches \cite{Ziegler1977,Filatov1999}. Unlike linear-response time-dependent density-functional theory (LR-TDDFT) \cite{Marques2006,Ullrich2012}, which determines excitation energies from the response of a ground-state reference, the $\Delta$SCF approach allows state-specific orbital relaxation. Such relaxation can be particularly beneficial for excitations involving substantial redistribution of the electron density, including charge-transfer states \cite{Selenius2024}.

To assess Skala's performance across different excitation classes and in the prediction of singlet-triplet gaps, we consider three benchmark subsets. QUEST1 \cite{Loos2018} comprises singlet and triplet excitations in small molecules. The ST-gap subset \cite{Loos2023} comprises heptazine, cyclazine, and related compounds exhibiting inverted singlet-triplet ordering, which standard adiabatic LR-TDDFT approximations generally fail to reproduce \cite{SanchoGarcia2022}. QUEST6 \cite{Loos2021} covers intramolecular charge-transfer excitations, which are challenging for LR-TDDFT with semilocal and conventional global-hybrid functionals, although range-separated hybrids can substantially improve their description.

We compare Skala with three conventional XC functionals representing successive rungs of Jacob's ladder: the GGA PBE \cite{Perdew1996}, the meta-GGA r$^2$SCAN \cite{Furness2020}, and the hybrid PBE0 \cite{Perdew1996a,Ernzerhof1999}. Open-shell excited states are treated using spin-unrestricted KS (UKS), with spin purification for open-shell singlets, and restricted open-shell KS (ROKS) \cite{Filatov1999,Kowalczyk2013}, allowing us to examine the dependence of excitation energies and singlet–triplet gaps on the open-shell treatment. This distinction is particularly relevant for inverted singlet–triplet gaps. Valverde et al. \cite{Valverde2025} attributed inverted singlet–triplet gaps obtained within UKS to differences in orbital relaxation and spin polarization between singlet and triplet states. In contrast, their ROKS calculations yielded positive gaps with the tested conventional hybrid functionals, whereas double hybrids, which include perturbative correlation, recovered inverted ordering for some systems. Skala's learned nonlocal XC dependence may alter the relative energies of mixed-spin and high-spin configurations. We therefore investigate whether Skala can reproduce inverted singlet–triplet ordering within ROKS.

\section{Theoretical background \label{sec:theory}}

In this work, we consider single-electron excitations from closed-shell ground states. Within the $\Delta$SCF framework, vertical excitation energies are obtained from differences between ground- and excited-state total energies evaluated at a common nuclear geometry,
\begin{equation}
  \Delta E_I=E_I-E_0,
  \label{eq:vertical_excitation}
\end{equation}
where $E_0$ and $E_I$ denote the ground- and excited-state total energies, respectively. The ground- and excited-state orbitals are optimized separately, with spin configurations and orbital occupations selected to target the desired electronic states. Depending on the targeted state, excited-state orbital optimization may involve saddle points rather than minima of the total energy. Consequently, appropriate state-targeting procedures are often required to avoid variational collapse to lower-energy solutions \cite{Gilbert2008,Barca2018,Hait2020,Herbert2023,Restaino2026}.

A common approach to state-specific excited-state calculations employs the spin-unrestricted Kohn-Sham (UKS) formalism. In this formalism, the auxiliary KS wavefunction is represented by an unrestricted Slater determinant, constructed from two separate sets of spatial orbitals for $\alpha$- and $\beta$-spin electrons. Although this determinant has a well-defined spin projection $M_S=(N_\alpha-N_\beta)/2$, it is generally not an eigenfunction of the squared total spin operator $\hat S^2$. Consequently, the determinant contains contributions from states with different total spin quantum numbers $S$, giving rise to spin contamination \cite{Amos1961,Lowdin1964}. Strong spin contamination is often associated with reduced accuracy of calculated energies. Moreover, the absence of well-defined total spin complicates the assignment and physical interpretation of electronic states, particularly in systems with several nearly degenerate spin states.

The description of open-shell singlet excited states requires particular care. A single-electron promotion from a closed-shell configuration produces two singly occupied spatial orbitals. In the corresponding $M_S=0$ configuration, these orbitals are occupied by electrons with opposite spins. The resulting single-determinant wavefunction approximates an equal mixture of a singlet and a triplet state. Such a mixed-spin determinant exhibits substantial spin contamination, with $\langle\hat S^2\rangle\approx1$ rather than the singlet value of zero. The corresponding UKS total energy can be used directly as an approximation to the open-shell singlet excited-state energy. However, an \emph{a posteriori} spin-purification procedure is commonly employed to improve the accuracy of the singlet-energy estimate by combining the mixed-spin energy with the energy of the corresponding triplet configuration \cite{Ziegler1977} as
\begin{equation}
    E_{\mathrm S}^{\mathrm{UKS}}
    = 2E_{\mathrm M}^{\mathrm{UKS}}
      - E_{\mathrm T}^{\mathrm{UKS}}.
    \label{eq:spin_purified_uks}
\end{equation}
Here, $E_{\mathrm S}^{\mathrm{UKS}}$ denotes the spin-purified estimate of the open-shell singlet energy, while $E_{\mathrm M}^{\mathrm{UKS}}$ and $E_{\mathrm T}^{\mathrm{UKS}}$ are the energies of the separately optimized mixed-spin $(M_S=0)$ and triplet $(M_S=1)$ determinants, respectively. The singlet excitation energy is then obtained by inserting $E_{\mathrm S}^{\mathrm{UKS}}$ into Eq.~\eqref{eq:vertical_excitation}.

An alternative treatment is provided by the restricted open-shell Kohn-Sham (ROKS) formalism \cite{Filatov1998,Filatov1999}. For open-shell singlet excited states, ROKS directly optimizes a singlet-energy expression analogous to Eq.~\eqref{eq:spin_purified_uks} \cite{Frank1998,Kowalczyk2013},
\begin{equation}
    E_{\mathrm S}^{\mathrm{ROKS}}[\{\phi_p\}]
    = 2E_{\mathrm M}[\{\phi_p\}]
      - E_{\mathrm T}[\{\phi_p\}],
    \label{eq:roks_singlet_energy}
\end{equation}
where the mixed-spin and triplet determinants are constructed from a common set of orthonormal spatial orbitals $\{\phi_p\}$ but differ in their occupation patterns in the $\alpha$- and $\beta$-spin channels. Unlike in the \emph{a posteriori} UKS procedure, these orbitals are optimized for the combined singlet energy rather than for the two determinant energies separately \cite{Herbert2023}. The triplet contribution is therefore evaluated with singlet-optimized orbitals, whereas the triplet excitation energies and singlet–triplet gaps reported here, unless stated otherwise, are obtained from separately optimized restricted open-shell triplet states.

The ROKS orbital equations involve occupation-dependent effective operators, making orbital optimization more involved than in conventional UKS calculations \cite{Kowalczyk2013}. These equations can be solved, for instance, using a generalized effective Fock matrix or through direct orbital optimization. Moreover, multiple solutions and artificial mixing of the open-shell orbitals can complicate convergence to the intended excited state \cite{Kowalczyk2013}. Appropriate initial orbitals and state-selection procedures are therefore important, together with checks of the converged state character and orbital stationarity. Alternatively, the singlet-energy expression in Eq.~\eqref{eq:roks_singlet_energy} can be optimized with respect to a common local multiplicative potential using the optimized effective potential (OEP) method \cite{Trushin2026}. This approach yields canonical KS orbital equations but is not employed in the present work.

The discussion above is limited to single-electron excitations from a spatially nondegenerate occupied orbital to a spatially nondegenerate unoccupied orbital. When either orbital belongs to a degenerate manifold, the description of the targeted electronic state may require symmetry-adapted combinations of orbitals or determinants. Such cases are not considered in the present work.

Finally, an important distinction relevant to the interpretation of our results concerns the lowest triplet and open-shell singlet excited states. For molecules with closed-shell singlet ground states, the lowest triplet can generally be obtained by minimizing the energy at a fixed spin projection $M_S=1$, provided that no higher-spin state has a lower energy \cite{Gunnarsson1976}. The corresponding excitation energy is therefore a difference between two lowest-energy solutions obtained at different fixed spin projections, without requiring explicit targeting of a non-Aufbau excited state. In contrast, open-shell singlet excited states have $M_S=0$, like the ground state, and require state-specific orbital occupations together with an appropriate treatment of spin coupling. Consequently, both open-shell singlet excitation energies and the corresponding singlet-triplet gaps depend on non-Aufbau excited-state solutions.

\section{Computational details \label{sec:comp_det}}

The calculations reported in the main text were performed with CP2K \cite{VandeVondele2005,Kuehne2020}, using the all-electron Gaussian and augmented plane-wave (GAPW) method \cite{Lippert1999,Krack2000} and the aug-cc-pVTZ orbital basis \cite{Dunning1989,Kendall1992,Woon1993}. Molecules were treated with nonperiodic boundary conditions and the Martyna-Tuckerman Poisson solver in cubic cells of side length 15~\AA, with larger cells used for selected QUEST6 systems. The plane-wave cutoff and relative cutoff were 400 and 50~Ry, respectively, and the standard SCF convergence threshold was \texttt{EPS\_SCF}$=10^{-5}$. PBE0 employed the CP2K resolution-of-the-identity Hartree–Fock exchange implementation \cite{Bussy2023}, using an automatically generated auxiliary basis set with the \texttt{MEDIUM} option. Skala~1.1 was evaluated through the molecular CP2K-GauXC interface \cite{Poeschel2026}, using the \texttt{skala-1.1-rev1} checkpoint. The atom-centered quadrature employed the GauXC \texttt{FINE} grid with \texttt{ROBUST} pruning and Mura-Knowles radial quadrature. PBE and PBE0 used native CP2K GAPW XC integration, while r$^2$SCAN employed the same framework with LibXC routines. No empirical dispersion correction was applied. The D3(BJ) correction associated with Skala \cite{Luise2026} depends only on the nuclear geometry and therefore contributes equally to all electronic states at a fixed geometry, canceling from the calculated vertical excitation energies and singlet-triplet gaps.

The benchmark subsets comprise 12 molecules from QUEST1 \cite{Loos2018}, 8 molecules from ST-gap \cite{Loos2023}, and 11 systems from QUEST6 \cite{Loos2021}. The subsets were selected to focus on single-electron excitations between nondegenerate spatial orbitals, avoiding complications associated with spatial symmetry breaking. Most of the targeted excitations were accessible through HOMO-LUMO promotions, facilitating consistent state initialization across XC functionals. The reference geometries and theoretical best estimates (TBEs) were used without further modification.

Ground-state orbitals were optimized within the restricted KS formalism. For QUEST1 and ST-gap, all targeted excitations involved highest occupied molecular orbital (HOMO) to lowest unoccupied molecular orbital (LUMO) promotions. In UKS, mixed-spin states were initialized by promoting a $\beta$-spin electron from the HOMO to the LUMO of the corresponding ground-state calculation. The orbitals were optimized using diagonalization with SCF mixing and DIIS, while the initial maximum overlap method (IMOM) \cite{Gilbert2008,Barca2018} was employed to maintain the targeted orbital occupations relative to a fixed initial reference. In ROKS, singlet energies were obtained by directly optimizing Eq.~\eqref{eq:roks_singlet_energy} using CP2K's low-spin ROKS implementation, employing orbital transformation (OT) with the DIIS minimizer. Unlike the diagonalization-based UKS calculations, ROKS singlet optimization employs only the explicitly represented occupied orbitals and does not use MOM-based occupation tracking. UKS triplet states were optimized separately by diagonalization with IMOM tracking from the first SCF iteration for all three datasets. The high-spin restricted open-shell triplets used in the QUEST1 and ST-gap ROKS calculations were optimized separately by diagonalization without MOM-based occupation tracking. Open-shell singlet energies within UKS were calculated using the spin-purification expression in Eq.~\eqref{eq:spin_purified_uks}. For QUEST6, only UKS results are reported because ROKS calculations encountered multiple solutions and mixing of open-shell orbitals. Some QUEST6 excitations required alternative orbital promotions or initial guesses from converged excited-state solutions.

To assess basis-set dependence, additional aug-cc-pVQZ calculations were performed. Independent calculations using a PySCF-based implementation \cite{Sun2026} were also carried out to cross-check the CP2K results. The Supplementary Material (SM) presents these comparisons and provides benchmark reference data and details of excited-state optimization procedures.

\section{Results \label{sec:results}}

\begin{table*}[t!]
\caption{\label{tab:mae_summary}
Mean absolute errors (MAEs, in eV) of singlet ($S$) and triplet ($T$) excitation energies and singlet-triplet gaps ($\Delta E_{\mathrm{ST}}$) obtained with $\Delta$SCF-UKS and $\Delta$SCF-ROKS for the QUEST1, ST-gap, and QUEST6 subsets. Open-shell singlet energies within UKS are obtained using spin purification. The lowest MAE for each subset, property, and treatment is highlighted in bold. For QUEST6, only UKS singlet excitation-energy MAEs are reported.
}
\begin{ruledtabular}
\begin{tabular}{llcccccc}
 & & \multicolumn{3}{c}{$\Delta$SCF-UKS} & \multicolumn{3}{c}{$\Delta$SCF-ROKS} \\
\cline{3-5}\cline{6-8}
Dataset & XC functional & $S$ & $T$ & $\Delta E_{\mathrm{ST}}$ & $S$ & $T$ & $\Delta E_{\mathrm{ST}}$ \\
\hline
QUEST1 & PBE & 0.507 & 0.255 & 0.252 & 0.485 & 0.198 & 0.307 \\
 & r$^2$SCAN & 0.482 & 0.435 & 0.103 & 0.438 & 0.298 & 0.166 \\
 & PBE0 & 0.411 & 0.358 & \textbf{0.085} & 0.389 & 0.258 & \textbf{0.147} \\
 & Skala & \textbf{0.170} & \textbf{0.060} & 0.183 & \textbf{0.275} & \textbf{0.128} & 0.272 \\
\hline
ST-gap & PBE & 0.130 & 0.153 & 0.113 & \textbf{0.158} & 0.149 & 0.232 \\
 & r$^2$SCAN & 0.275 & 0.135 & 0.140 & 0.182 & 0.085 & 0.241 \\
 & PBE0 & 0.256 & \textbf{0.078} & 0.178 & 0.267 & \textbf{0.031} & 0.278 \\
 & Skala & \textbf{0.125} & 0.110 & \textbf{0.050} & 0.257 & 0.170 & \textbf{0.087} \\
\hline
QUEST6 & PBE & 0.775 & - & - & - & - & - \\
 & r$^2$SCAN & 0.599 & - & - & - & - & - \\
 & PBE0 & \textbf{0.335} & - & - & - & - & - \\
 & Skala & 0.422 & - & - & - & - & - \\
\end{tabular}
\end{ruledtabular}
\end{table*}

\begin{figure*}[!t]
    \centering
    \includegraphics[width=0.65\linewidth]{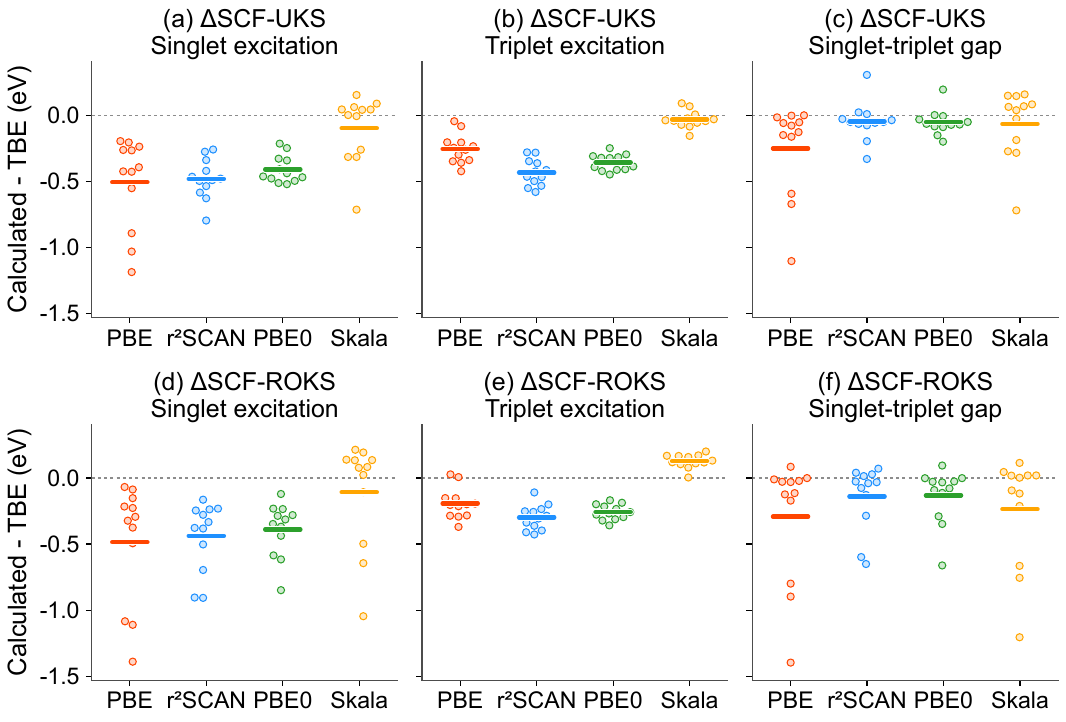}
    \caption{Signed errors relative to theoretical best estimates for singlet excitation energies (a,d), triplet excitation energies (b,e), and singlet-triplet gaps, $E_{\mathrm{S}}-E_{\mathrm{T}}$ (c,f), for the 12-molecule QUEST1 subset. The upper and lower rows show $\Delta$SCF-UKS and $\Delta$SCF-ROKS results, respectively, for PBE, r$^2$SCAN, PBE0, and Skala. Each circle represents one molecule; colored horizontal bars indicate mean signed errors. Errors are defined as calculated values minus reference values, so positive values indicate overestimation.}
    \label{fig:quest1_signed_errors}
\end{figure*}

We first examine the selected singlet and triplet excitation energies and the corresponding singlet-triplet gaps for the 12-molecule QUEST1 subset. For ethylene and cyclopropene, the selected singlets lie above $S_1$, so the reported gaps refer to the targeted $\pi\to\pi^*$ singlet and triplet states. Table~\ref{tab:mae_summary} summarizes the mean absolute errors (MAEs) relative to the theoretical best estimates for PBE, r$^2$SCAN, PBE0, and Skala, using $\Delta$SCF-UKS and $\Delta$SCF-ROKS calculations.

Skala yields the lowest MAEs for both singlet and triplet excitation energies with either treatment. For $\Delta$SCF-UKS, its singlet MAE is 0.170~eV, compared with 0.411-0.507~eV for the conventional functionals. The improvement is particularly pronounced for triplet excitations, for which Skala achieves an MAE of 0.060~eV, compared with 0.255-0.435~eV for the other functionals. Skala also retains the lowest excitation-energy errors with ROKS, with singlet and triplet MAEs of 0.275 and 0.128~eV, respectively. However, both errors increase relative to the $\Delta$SCF-UKS results, whereas the conventional functionals generally exhibit lower excitation-energy MAEs with ROKS. The particularly small triplet errors may partly reflect the fact that lowest-triplet excitation energies involve differences between lowest-energy solutions at different spin projections, making them closer in character to the ground-state energetics that dominate Skala's training data. In contrast, open-shell singlet excitation energies and the corresponding singlet-triplet gaps additionally probe non-Aufbau excited-state solutions that were not explicitly targeted during training.

The signed-error distributions for QUEST1 in Fig.~\ref{fig:quest1_signed_errors} further illustrate the differences in accuracy between singlet and triplet excitation energies. With $\Delta$SCF-UKS, the conventional functionals systematically underestimate both singlet and triplet excitation energies. Skala substantially reduces this negative bias, particularly for triplet excitations, whose errors are concentrated near zero. Its largest singlet underestimations occur for ethylene (-0.715~eV), cyclopropene (-0.316~eV), and the streptocyanine-C1 cation (-0.316~eV). These deviations become more pronounced with ROKS, reaching -1.046, -0.645, and -0.498~eV, respectively. The remaining ROKS singlet errors are predominantly positive, resulting in a relatively small mean signed error despite the substantial deviations for these three molecules. In contrast, Skala's ROKS triplet errors are positive throughout the subset.

The ranking changes when considering singlet-triplet gaps. PBE0 gives the lowest gap MAEs with both $\Delta$SCF-UKS and $\Delta$SCF-ROKS, at 0.085 and 0.147~eV, respectively, followed by r$^2$SCAN at 0.103 and 0.166~eV. The corresponding Skala errors are 0.183 and 0.272~eV. This behavior illustrates the importance of error cancellation: the gap error is the difference between the singlet and triplet excitation-energy errors. The systematic underestimation of both excitation energies by PBE0 and r$^2$SCAN permits favorable error cancellation, whereas for Skala with ROKS, the negative singlet errors of the three molecules identified above combine with positive triplet errors, amplifying the gap underestimation. These three cases account for approximately 80\% of the total absolute ROKS gap error.

To assess the sensitivity of the results to these three outliers, we recomputed the MAEs after excluding them from all functional and method comparisons, leaving a common nine-molecule subset. Skala retains the lowest MAEs for singlet and triplet excitation energies, reaching 0.078 and 0.071~eV with $\Delta$SCF-UKS and 0.123 and 0.122~eV with $\Delta$SCF-ROKS, respectively. Its gap MAEs decrease from 0.183 to 0.102~eV with UKS and from 0.272 to 0.071~eV with ROKS. Thus, excluding these three molecules reverses the relative accuracy of UKS and ROKS for Skala's singlet-triplet gaps, with ROKS now yielding the smaller MAE. Nevertheless, all three conventional functionals retain lower gap MAEs than Skala on the reduced subset.

\begin{figure*}[!t]
    \centering
    \includegraphics[width=0.65\linewidth]{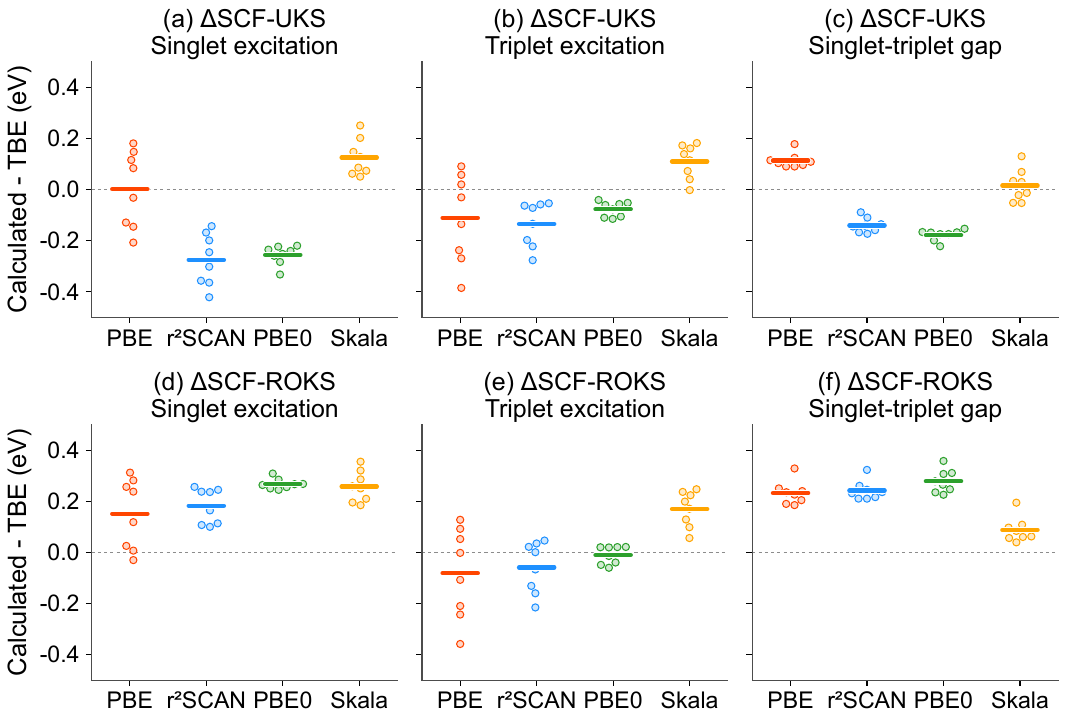}
    \caption{Signed errors relative to theoretical best estimates for singlet excitation energies (a,d), triplet excitation energies (b,e), and singlet-triplet gaps, $\Delta E_{\mathrm{ST}}=E_{\mathrm{S}}-E_{\mathrm{T}}$ (c,f), for the eight-molecule ST-gap subset. The upper and lower rows show $\Delta$SCF-UKS and $\Delta$SCF-ROKS results, respectively, for PBE, r$^2$SCAN, PBE0, and Skala. Each circle represents one molecule; colored horizontal bars indicate mean signed errors. Errors are defined as calculated values minus reference values, so positive values indicate overestimation.}
    \label{fig:st_gap_signed_errors}
\end{figure*}

\begin{figure*}[!t]
    \centering
    \includegraphics[width=0.625\linewidth]{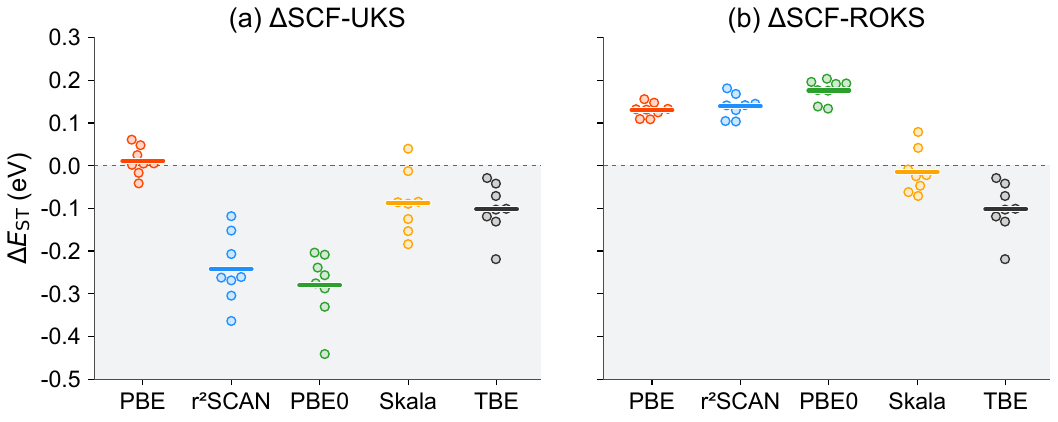}
    \caption{Singlet-triplet energy gaps, $\Delta E_{\mathrm{ST}} = E_{\mathrm{S}} - E_{\mathrm{T}}$, for the eight-molecule ST-gap subset, calculated using (a) $\Delta$SCF-UKS and (b) $\Delta$SCF-ROKS. Circles represent individual molecules, and short horizontal bars indicate mean gaps. TBEs are shown in gray as the rightmost group in each panel. The dashed line marks zero gap; the shaded region indicates inverted ordering, with the singlet below the triplet.}
    \label{fig:st_gap_gap_distributions}
\end{figure*}

\begin{figure}[!t]
    \centering
    \includegraphics[width=\linewidth]{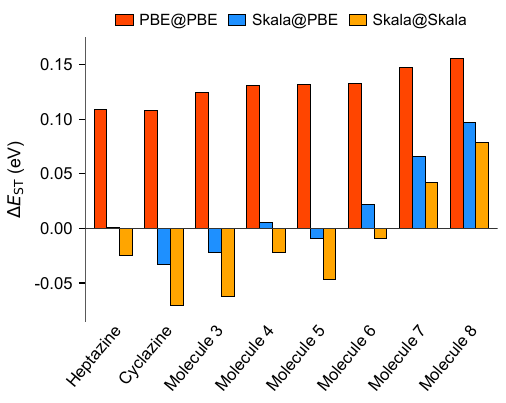}
    \caption{Singlet-triplet gaps for the eight molecules of the ST-gap subset calculated using ROKS. Results are shown for PBE with optimized PBE orbitals (PBE@PBE), Skala evaluated on fixed PBE-optimized orbitals (Skala@PBE), and Skala with fully optimized orbitals (Skala@Skala). Each gap combines the ROKS singlet energy on donor singlet orbitals with the triplet energy on separately optimized donor triplet orbitals. Negative gaps indicate inverted singlet-triplet ordering, which is obtained for zero, three, and six molecules, respectively. The molecular labels follow the nomenclature of the original ST-gap benchmark dataset.}
\label{fig:st_gap_roks_analysis}
\end{figure}

We next consider the eight-molecule ST-gap subset comprising heptazine, cyclazine, and related compounds. All reference singlet-triplet gaps in this subset are negative, providing a targeted test of the ability of the functionals to reproduce inverted excited-state ordering. Table~\ref{tab:mae_summary} summarizes the mean absolute errors (MAEs) of the singlet and triplet excitation energies and singlet-triplet gaps, while Fig.~\ref{fig:st_gap_signed_errors} presents the corresponding signed errors. Figure~\ref{fig:st_gap_gap_distributions} compares the calculated singlet-triplet gaps with the theoretical best estimates.

Skala yields the lowest singlet-triplet gap MAE with both computational treatments. For $\Delta$SCF-UKS, its MAE is 0.050~eV, compared with 0.113, 0.140, and 0.178~eV for PBE, r$^2$SCAN, and PBE0, respectively.
With the ROKS-based treatment, the Skala gap MAE increases to 0.087~eV but remains substantially below those of the conventional functionals, which range from 0.232 to 0.278~eV. This ranking differs from that obtained for QUEST1, where PBE0 and r$^2$SCAN provide more accurate singlet-triplet gaps than Skala.

The advantage of Skala for the gaps is not accompanied by uniformly smaller errors in the individual excitation energies. For $\Delta$SCF-UKS, its singlet MAE of 0.125~eV is comparable to the PBE value of 0.130~eV, while its triplet MAE of 0.110~eV exceeds the PBE0 value of 0.078~eV. With the ROKS-based treatment, Skala gives singlet and triplet MAEs of 0.257 and 0.170~eV, respectively. The former exceeds the corresponding PBE and r$^2$SCAN errors, while the latter is the largest among the four functionals. As shown in Fig.~\ref{fig:st_gap_signed_errors}, Skala exhibits predominantly positive excitation-energy errors, with systematic overestimation of both singlet and triplet energies in the ROKS-based calculations. Partial cancellation of these errors leads to the smaller error in their difference. The accuracy of the gaps must therefore be distinguished from the accuracy of the individual excitation energies.

Figure~\ref{fig:st_gap_gap_distributions} distinguishes the accuracy of the gap magnitudes from the prediction of their signs. With spin-purified $\Delta$SCF-UKS, r$^2$SCAN and PBE0 predict inverted ordering for all eight molecules but systematically overestimate the magnitude of the inversion. Skala reproduces the inverted ordering for seven of eight molecules, with gap values generally closer to the reference estimates, whereas PBE produces gaps clustered around zero. The ROKS-based calculations show a different pattern: PBE, r$^2$SCAN, and PBE0 predict positive gaps for all eight molecules, while Skala predicts negative gaps for compounds 1-6 and positive gaps for compounds 7 and 8. Notably, although the reference gaps for compounds 7 and 8 are negative ($-0.042$ and $-0.029$~eV, respectively), their magnitudes were considered too small to establish the inverted ordering definitively in the original benchmark \cite{Loos2023}. Thus, Skala reproduces the inverted ordering for all six molecules in the selected subset whose negative reference gaps were considered definitive. However, the predicted gap magnitudes are not necessarily accurate, as illustrated by heptazine, for which Skala/ROKS yields -0.025~eV compared with the reference value of -0.219~eV.

To examine Skala's distinctive behavior within ROKS, we investigated the dependence of the singlet-triplet gaps on the XC functional and the state-specific orbital sets. Specifically, we evaluated the ROKS singlet and triplet energies with Skala using their respective PBE-optimized orbital sets, without further optimization. As shown in Fig.~\ref{fig:st_gap_roks_analysis}, replacing PBE with Skala at fixed PBE-optimized orbitals reduces the mean gap by 0.114~eV and yields negative gaps for three molecules. Reoptimizing both the singlet and triplet orbital sets with Skala produces a further net reduction of 0.030~eV, increasing the number of negative gaps to six. These changes describe the chosen comparison path and do not constitute a unique decomposition of functional and orbital effects. The molecular results and an alternative comparison path are provided in the SM.

We next compare singlet-triplet splittings evaluated on the same set of ROKS singlet-optimized orbitals with gaps obtained after separate triplet optimization. In Eq.~\eqref{eq:roks_singlet_energy}, the mixed-spin and triplet energy components are evaluated using the same spatial orbitals optimized for the ROKS open-shell singlet. The two configurations occupy the same spatial orbitals and therefore have identical total electron densities and noninteracting kinetic energies, but different spin densities. Consequently, their non-XC energy contributions cancel, leaving the common-orbital splitting determined entirely by the difference between their XC energies. Using the triplet component instead of the separately optimized triplet energy yields negative common-orbital splittings for seven of eight ST-gap molecules with Skala, with a mean of -0.066~eV. In contrast, PBE, r$^2$SCAN, and PBE0 yield positive common-orbital splittings for all eight molecules. Separate optimization of the triplet orbitals lowers the triplet energies and increases the Skala gaps by 0.052~eV on average, leaving six negative gaps with a mean of -0.014~eV. Thus, Skala yields negative singlet-triplet splittings before separate triplet optimization, which opposes rather than induces the inverted ordering. Together, these comparisons indicate that Skala's distinctive ROKS behavior is associated with its treatment of spin-dependent XC energetics rather than separate triplet relaxation. The molecular results for all four functionals and the definitions of both gap evaluations are provided in the SM.

The comparison between the UKS and ROKS treatments reveals a systematic sensitivity of the predicted singlet-triplet gaps to the open-shell treatment. For all four functionals, ROKS yields larger gap MAEs than spin-purified $\Delta$SCF-UKS. Skala retains the lowest gap MAEs among the tested functionals under both treatments, while the differences in the MAEs for individual singlet and triplet excitation energies further highlight the importance of the chosen open-shell treatment.

\begin{figure}[!t]
    \centering
    \includegraphics[width=\linewidth]{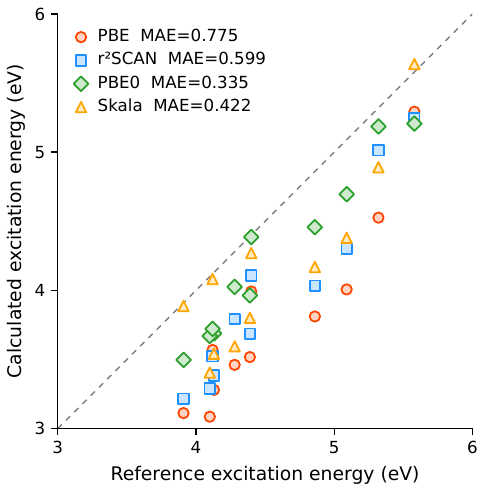}
    \caption{Spin-purified $\Delta$SCF-UKS singlet excitation energies calculated with PBE, r$^2$SCAN, PBE0, and Skala for the 11-system QUEST6 subset, plotted against theoretical best estimates (TBEs). The dashed line indicates perfect agreement, and the legend reports the mean absolute errors (MAEs) in eV.}
    \label{fig:quest6_scatter}
\end{figure}

Finally, we consider singlet intramolecular charge-transfer (CT) excitations in the 11-system QUEST6 subset. Such excitations are known to be challenging for LR-TDDFT with semilocal and conventional global-hybrid XC functionals \cite{Loos2021,Selenius2024}. Although orbital-optimized $\Delta$SCF allows state-specific orbital relaxation, previous benchmarks have also found larger CT excitation-energy errors than for local valence and Rydberg excitations when using semilocal functionals \cite{Trushin2026}. As summarized in Table~\ref{tab:mae_summary}, PBE0 yields the lowest MAE of 0.335~eV for spin-purified $\Delta$SCF-UKS excitation energies, followed by Skala at 0.422~eV, r$^2$SCAN at 0.599~eV, and PBE at 0.775~eV. The scatter plot in Fig.~\ref{fig:quest6_scatter} shows that Skala has smaller absolute errors than PBE0 for three of the eleven excitations (phthalazine, twisted DMABN, and twisted PP), whereas PBE0 is more accurate for the other eight. PBE0 underestimates all eleven reference excitation energies, while Skala underestimates ten of them. The QUEST6 singlet MAEs are generally larger than those obtained with UKS for QUEST1 and ST-gap, consistent with the known challenges of describing CT excitations. Thus, although Skala outperforms PBE and r$^2$SCAN for QUEST6, PBE0 remains more accurate on average, highlighting the dependence of Skala's relative performance on the excitation class and electronic property considered.

\section{Conclusions \label{sec:conclusions}}

We have assessed the performance of the machine-learned Skala XC functional for molecular vertical excitation energies and singlet-triplet gaps within the orbital-optimized $\Delta$SCF framework. We compared Skala with PBE, r$^2$SCAN, and PBE0 using selected systems from the QUEST1, ST-gap, and QUEST6 datasets. Calculations were performed using UKS, including spin purification for open-shell singlets, and ROKS, with the latter applied only to QUEST1 and ST-gap.

For QUEST1, Skala yields the lowest mean absolute errors for singlet and triplet excitation energies with both UKS and ROKS, reaching MAEs of 0.170 and 0.060~eV, respectively, with UKS. However, PBE0 and r$^2$SCAN provide more accurate singlet-triplet gaps due to more favorable error cancellation, highlighting the distinction between the accuracy of individual excitation energies and that of their differences. For ST-gap, Skala instead achieves the lowest gap MAEs of 0.050 and 0.087~eV with UKS and ROKS, respectively. Notably, within ROKS, Skala predicts inverted ordering for all six molecules in the selected subset whose negative reference gaps were considered definitive \cite{Loos2023}, whereas all three conventional functionals yield positive gaps throughout the subset.

The ROKS analysis provides further insight into the origin of these inverted gaps. Evaluating the mixed-spin and triplet energy components on the same orbitals optimized for the ROKS open-shell singlet yields negative splittings for seven of eight molecules with Skala, whereas all three conventional functionals yield positive splittings on their respective singlet-optimized orbital sets. Since the non-XC contributions cancel on the common orbital set, these splittings are determined entirely by differences in the XC energies of the mixed-spin and high-spin configurations. Separate optimization of the triplet increases the Skala gaps by 0.052~eV on average, but six negative gaps remain. Thus, the inverted ordering is associated with Skala's distinct treatment of spin-dependent XC energetics and persists despite the opposing effect of separate triplet relaxation.

For QUEST6, where only UKS calculations were considered, Skala yields a singlet excitation-energy MAE of 0.422~eV, outperforming PBE and r$^2$SCAN but remaining less accurate than PBE0 (0.335~eV). Across QUEST1 and ST-gap, Skala generally achieves lower MAEs with UKS than with ROKS, highlighting the importance of the open-shell treatment. The particularly accurate lowest-triplet excitation energies obtained with Skala for QUEST1 may partly reflect their similarity to the ground-state energetics represented in its training data, whereas open-shell singlet excitations and the corresponding gaps provide a more direct test of transferability to non-Aufbau electronic states. Overall, these results provide evidence of Skala's transferability beyond ground-state energetics to molecular excited states without excited-state-specific retraining.

\section*{Supplementary Material}

The Supplementary Material (SM) provides details on the selection of systems and electronic states, basis-set dependence, comparisons with a PySCF-based implementation, and ROKS analyses of functional substitution, state-specific orbital reoptimization, and separate triplet optimization for the ST-gap subset.

\section*{Acknowledgments}

Part of the research was funded by the Deutsche Forschungsgemeinschaft (DFG, German Research Foundation), project number 417590517/CRC~1415.

\section*{Author declarations}
\subsection*{Conflict of Interest}

The authors have no conflicts to disclose.

\section*{Data Availability}

The data supporting this study are archived on Zenodo~\cite{Zenodo_repo}, including CP2K calculation inputs and outputs, molecular geometries, calculated excitation energies and singlet--triplet gaps, reference values, and analysis scripts for reproducing the manuscript tables and figures. Results from the PySCF-based calculations are provided as CSV files; the original PySCF inputs, outputs, and custom implementation are not included. The analysis scripts are also available on GitHub~\cite{GitHub_repo}.

\bibliography{bibliography}

\clearpage
\onecolumngrid
\setcounter{section}{0}
\setcounter{table}{0}
\setcounter{equation}{0}
\setcounter{figure}{0}
\renewcommand{\thesection}{S\arabic{section}}
\renewcommand{\thetable}{S\arabic{table}}
\renewcommand{\theequation}{S\arabic{equation}}
\renewcommand{\thefigure}{S\arabic{figure}}
\renewcommand{\theHsection}{supplement.\arabic{section}}
\renewcommand{\theHtable}{supplement.\arabic{table}}
\renewcommand{\theHequation}{supplement.\arabic{equation}}
\renewcommand{\theHfigure}{supplement.\arabic{figure}}
\input{si/supplementary_body}

\end{document}

%% file: si/supplementary_body.tex
\providecommand{\sipending}{\textit{TBD}}

\begin{center}
{\large Supplementary Material for: Beyond the Ground State: Assessing the Machine-Learned Skala Functional in $\Delta$SCF Calculations of Molecular Excitation Energies}\\[1ex]
Egor Trushin and Thomas D. K\"uhne\\[0.5ex]
Author affiliations are given in the main manuscript.
\end{center}

The Supplementary Material provides details on the selection of systems and electronic states, numerical comparisons using a larger orbital basis and a PySCF-based implementation, and ROKS analyses of functional substitution, state-specific orbital reoptimization, and separate triplet optimization for the ST-gap subset. The computational methods and numerical settings are described in the main text. CP2K calculation inputs and outputs, molecular geometries, state-resolved energy tables, PySCF-based numerical results and calculation details, and analysis scripts are available in the data repository archived on Zenodo~\cite{Zenodo_repo}. The scripts for reproducing the manuscript tables and figures are also available on GitHub~\cite{GitHub_repo}. The PySCF-based comparisons can be reproduced from the provided CSV files; the original PySCF inputs and outputs and the custom implementation are not included.

\section{Selection of systems and excitations}
\label{sec:si_selection}

The study uses 12 molecules from QUEST1~\cite{Loos2018}, 8 molecules from the ST-gap benchmark~\cite{Loos2023}, and 11 systems from QUEST6~\cite{Loos2021}. The subsets were selected to focus on single-electron excitations between nondegenerate spatial orbitals, avoiding complications associated with spatial symmetry breaking and simplifying the identification of targeted electronic states. Preference was also given to excitations accessible through HOMO--LUMO promotions to facilitate consistent state initialization across XC functionals. Compounds~9 and~10 of the ST-gap benchmark were excluded because targeting their excitations required HOMO$\to$LUMO+1 promotions for some functionals. QUEST1 and ST-gap were evaluated with UKS and ROKS, whereas QUEST6 was evaluated with UKS only.

Tables~\ref{tab:si_states_quest1}--\ref{tab:si_states_quest6} identify the selected reference states and theoretical best estimates (TBEs) used in the error analysis. Molecular geometries were taken from the corresponding reference datasets. Symmetry labels follow the reference publications, while ST-gap systems retain the original compound numbering. For QUEST1, all 12 selected triplets correspond to the lowest triplet excited states ($T_1$), whereas the selected singlets correspond to the lowest singlet excited states ($S_1$) for ten molecules. The exceptions are ethylene and cyclopropene, where the targeted $\pi\to\pi^*$ singlet states lie above lower-energy $\pi\to3s$ and $\sigma\to\pi^*$ singlet states, respectively. Consequently, the reported singlet--triplet gaps for these two molecules refer to the selected $\pi\to\pi^*$ singlet and triplet states rather than the $S_1$--$T_1$ energy differences. For QUEST6, the $S_n$ labels indicate the reference LR-CCSD/cc-pVTZ ordering, which need not be preserved by other methods. These labels identify the benchmark targets and are distinct from the orbital indices used to initialize the calculations.

HOMO--LUMO promotions were used for the selected QUEST1 and ST-gap systems. In QUEST6, twisted PP required a HOMO$\to$LUMO+1 promotion with PBE, r$^2$SCAN, and Skala. The PBE0 mixed-spin calculations for dimethylaniline, twisted DMABN, and twisted PP were instead initialized from the corresponding converged PBE mixed-spin determinants. Only singlet excitation energies were included in the QUEST6 benchmark statistics. The corresponding high-spin triplet states were optimized separately for the UKS spin-purification correction but were not independently benchmarked, as QUEST6 does not provide reference triplet excitation energies.

The converged excited-state solutions were examined using orbital-based diagnostics to assess retention of the targeted electronic configurations and identify possible convergence to unintended states. For UKS, these checks included orbital correspondence with the initial excitation, spin contamination, and consistency between the mixed-spin and separately optimized triplet states. For ROKS, the occupied and active orbital spaces were analyzed, together with their correspondence to the intended singlet configuration and separately optimized triplet state. Orbital character and spatial symmetry were additionally examined in selected cases, particularly when alternative initial excitations were required. These checks were used to assess the converged electronic configurations rather than relying solely on numerical convergence or agreement with reference excitation energies.

As an additional consistency check, we compared our PBE excitation energies with previously published orbital-optimized DFT results. For seven QUEST1 molecules, the singlet and triplet excitation energies agree closely with those reported by Ivanov et al.~\cite{Ivanov2021}, with mean and maximum absolute differences of approximately 0.015 and 0.025~eV, respectively. For all 11 QUEST6 systems, comparison with the results of Selenius et al.~\cite{Selenius2024} yields corresponding differences of approximately 0.017 and 0.058~eV. This agreement provides further support for the consistency of the targeted electronic states across different implementations and orbital-optimization procedures.

\begin{table}[htbp]
\centering
\caption{Selected QUEST1 targets and corrected TBEs (eV) from Table~6 of Ref.~\onlinecite{Loos2018}. Character refers to the singlet; V and R denote valence and Rydberg excitations.}
\label{tab:si_states_quest1}
\small
\setlength{\tabcolsep}{6pt}
\begin{tabular}{@{}lcccrr@{}}
\hline
System & Singlet & Triplet & Singlet character & $\Delta E_{\mathrm{S}}^{\mathrm{TBE}}$ & $\Delta E_{\mathrm{T}}^{\mathrm{TBE}}$ \\
\hline
Acetaldehyde & ${}^{1}A^{\prime\prime}$ & ${}^{3}A^{\prime\prime}$ & V; $n\to\pi^*$ & 4.31 & 3.98 \\
Ammonia & ${}^{1}A_2$ & ${}^{3}A_2$ & R; $n\to3s$ & 6.66 & 6.37 \\
Cyclopropene & ${}^{1}B_2$ & ${}^{3}B_2$ & V; $\pi\to\pi^*$ & 6.78 & 4.38 \\
Diazomethane & ${}^{1}A_2$ & ${}^{3}A_2$ & V; $\pi\to\pi^*$ & 3.13 & 2.80 \\
Ethylene & ${}^{1}B_{1u}$ & ${}^{3}B_{1u}$ & V; $\pi\to\pi^*$ & 7.92 & 4.54 \\
Formaldehyde & ${}^{1}A_2$ & ${}^{3}A_2$ & V; $n\to\pi^*$ & 3.97 & 3.58 \\
Ketene & ${}^{1}A_2$ & ${}^{3}A_2$ & V; $\pi\to\pi^*$ & 3.86 & 3.77 \\
Methanimine & ${}^{1}A^{\prime\prime}$ & ${}^{3}A^{\prime\prime}$ & V; $n\to\pi^*$ & 5.21 & 4.64 \\
Nitrosomethane & ${}^{1}A^{\prime\prime}$ & ${}^{3}A^{\prime\prime}$ & V; $n\to\pi^*$ & 1.95 & 1.16 \\
Streptocyanine-C1 ($q=+1$) & ${}^{1}B_2$ & ${}^{3}B_2$ & V; $\pi\to\pi^*$ & 7.12 & 5.52 \\
Thioformaldehyde & ${}^{1}A_2$ & ${}^{3}A_2$ & V; $n\to\pi^*$ & 2.20 & 1.94 \\
Water & ${}^{1}B_1$ & ${}^{3}B_1$ & R; $n\to3s$ & 7.70 & 7.33 \\
\hline\hline
\end{tabular}
\end{table}

\begin{table}[htbp]
\centering
\caption{Selected ST-gap systems and vertical TBEs (eV) for $S_1$ and $T_1$ from Table~S1 in the SM of Ref.~\onlinecite{Loos2023}. Compound numbering follows that reference; $\Delta E_{\mathrm{ST}}=\Delta E_{\mathrm{S}}-\Delta E_{\mathrm{T}}$. The original benchmark considered the small negative gaps of compounds~7 and~8 insufficient to establish inverted ordering definitively.}
\label{tab:si_states_st_gap}
\small
\setlength{\tabcolsep}{6pt}
\begin{tabular}{@{}lrrr@{}}
\hline
System & $\Delta E_{\mathrm{S}}^{\mathrm{TBE}}$ & $\Delta E_{\mathrm{T}}^{\mathrm{TBE}}$ & $\Delta E_{\mathrm{ST}}^{\mathrm{TBE}}$ \\
\hline
Heptazine (1) & 2.717 & 2.936 & -0.219 \\
Cyclazine (2) & 0.979 & 1.110 & -0.131 \\
Molecule~3 & 1.562 & 1.663 & -0.101 \\
Molecule~4 & 2.177 & 2.296 & -0.119 \\
Molecule~5 & 2.127 & 2.230 & -0.103 \\
Molecule~6 & 0.833 & 0.904 & -0.071 \\
Molecule~7 & 0.693 & 0.735 & -0.042 \\
Molecule~8 & 0.554 & 0.583 & -0.029 \\
\hline\hline
\end{tabular}
\end{table}

\begin{table}[htbp]
\centering
\caption{Selected QUEST6 singlet CT targets and aug-cc-pVQZ TBEs (eV) from Table~1 of Ref.~\onlinecite{Loos2021}. The $S_n$ labels follow the LR-CCSD/cc-pVTZ ordering in its SM (Table~S1).}
\label{tab:si_states_quest6}
\small
\setlength{\tabcolsep}{6pt}
\begin{tabular}{@{}lccr@{}}
\hline
System & Reference singlet & Character & $\Delta E_{\mathrm{S}}^{\mathrm{TBE}}$ \\
\hline
Aminobenzonitrile & $S_{2}$ (${}^{1}A_1$) & $\pi\to\pi^*$ & 5.09 \\
Benzothiadiazole & $S_{2}$ (${}^{1}B_2$) & $\pi\to\pi^*$ & 4.28 \\
Dimethylaminobenzonitrile (DMABN) & $S_{2}$ (${}^{1}A_1$) & $\pi\to\pi^*$ & 4.86 \\
Dimethylaniline & $S_{1}$ (${}^{1}B_2$) & $\pi\to\pi^*$ & 4.40 \\
Nitroaniline & $S_{4}$ (${}^{1}A_1$) & $\pi\to\pi^*$ & 4.39 \\
Nitrodimethylaniline & $S_{2}$ (${}^{1}A_1$) & $\pi\to\pi^*$ & 4.13 \\
Nitropyridine $N$-oxide & $S_{4}$ (${}^{1}A_1$) & $\pi\to\pi^*$ & 4.10 \\
$N$-Phenylpyrrole (PP) & $S_{3}$ (${}^{1}B_2$) & $\pi\to\pi^*$ & 5.32 \\
Phthalazine & $S_{1}$ (${}^{1}A_2$) & $n\to\pi^*$ & 3.91 \\
Twisted DMABN & $S_{1}$ (${}^{1}A_2$) & $n\to\pi^*$ & 4.12 \\
Twisted PP & $S_{2}$ (${}^{1}B_2$) & $\pi\to\pi^*$ & 5.58 \\
\hline\hline
\end{tabular}
\end{table}

\clearpage
\section{Basis-set dependence within CP2K}
\label{sec:si_basis}

To assess basis-set dependence, we compared excitation energies and singlet--triplet gaps obtained with CP2K using the aug-cc-pVTZ and aug-cc-pVQZ basis sets. Table~\ref{tab:si_basis} summarizes the mean and maximum absolute differences between the two basis sets.

\begin{table}[htbp]
\centering
\caption{Mean absolute differences (MADs) and maximum absolute differences (MaxADs) between CP2K aug-cc-pVTZ and aug-cc-pVQZ excitation energies and singlet--triplet gaps (eV). Statistics include 12 molecules for QUEST1 and eight for ST-gap, where available. Dashes indicate unavailable aug-cc-pVQZ results for ST-gap PBE0. No aug-cc-pVQZ calculations were performed for QUEST6.}
\label{tab:si_basis}
\footnotesize
\setlength{\tabcolsep}{4pt}
\begin{tabular}{@{}llrrrrrr@{}}
\hline
Treatment & XC functional & \multicolumn{2}{c}{$\Delta E_{\mathrm{S}}$} & \multicolumn{2}{c}{$\Delta E_{\mathrm{T}}$} & \multicolumn{2}{c}{$\Delta E_{\mathrm{ST}}$} \\
\cline{3-4}\cline{5-6}\cline{7-8}
 & & MAD & MaxAD & MAD & MaxAD & MAD & MaxAD \\
\hline
\multicolumn{8}{l}{\textit{QUEST1}} \\
UKS & PBE & 0.004 & 0.012 & 0.003 & 0.011 & 0.002 & 0.005 \\
UKS & r$^2$SCAN & 0.007 & 0.014 & 0.006 & 0.011 & 0.003 & 0.009 \\
UKS & PBE0 & 0.004 & 0.012 & 0.003 & 0.011 & 0.002 & 0.005 \\
UKS & Skala & 0.015 & 0.035 & 0.010 & 0.028 & 0.013 & 0.053 \\
\noalign{\smallskip}
ROKS & PBE & 0.003 & 0.010 & 0.002 & 0.009 & 0.002 & 0.003 \\
ROKS & r$^2$SCAN & 0.004 & 0.010 & 0.002 & 0.006 & 0.003 & 0.010 \\
ROKS & PBE0 & 0.003 & 0.008 & 0.002 & 0.009 & 0.001 & 0.002 \\
ROKS & Skala & 0.008 & 0.023 & 0.006 & 0.012 & 0.011 & 0.019 \\
\hline
\multicolumn{8}{l}{\textit{ST-gap}} \\
UKS & PBE & 0.002 & 0.003 & 0.001 & 0.002 & 0.001 & 0.002 \\
UKS & r$^2$SCAN & 0.003 & 0.005 & 0.003 & 0.005 & 0.000 & 0.002 \\
UKS & PBE0 & -- & -- & -- & -- & -- & -- \\
UKS & Skala & 0.012 & 0.019 & 0.013 & 0.019 & 0.002 & 0.003 \\
\noalign{\smallskip}
ROKS & PBE & 0.001 & 0.001 & 0.000 & 0.001 & 0.001 & 0.001 \\
ROKS & r$^2$SCAN & 0.001 & 0.003 & 0.001 & 0.003 & 0.000 & 0.000 \\
ROKS & PBE0 & -- & -- & -- & -- & -- & -- \\
ROKS & Skala & 0.002 & 0.005 & 0.002 & 0.006 & 0.001 & 0.003 \\
\hline\hline
\end{tabular}
\end{table}

For QUEST1, Skala exhibits larger basis-set differences than the conventional functionals, particularly with UKS. Its maximum absolute differences reach 0.035 and 0.028~eV for singlet and triplet excitation energies, respectively, and 0.053~eV for singlet--triplet gaps. The corresponding gap differences are smaller with ROKS, with a maximum of 0.019~eV. In contrast, the maximum gap differences for PBE, r$^2$SCAN, and PBE0 do not exceed 0.010~eV with either treatment.

For ST-gap, all available calculations yield maximum singlet--triplet gap differences of at most 0.003~eV. Although the maximum differences in individual excitation energies reach 0.019~eV for Skala with UKS, the corresponding gap differences remain small, indicating favorable cancellation of basis-set effects. Importantly, the number of inverted singlet--triplet gaps remains unchanged upon increasing the basis set from aug-cc-pVTZ to aug-cc-pVQZ. Overall, despite somewhat larger basis-set differences for Skala on QUEST1, these results indicate that aug-cc-pVTZ provides sufficiently converged excitation energies and singlet--triplet gaps for the QUEST1 and ST-gap systems included in the comparison, making it a suitable choice for practical calculations.

\clearpage
\section{Comparison with a PySCF-based implementation}
\label{sec:si_pyscf}

Here we compare singlet and triplet excitations and singlet--triplet gaps obtained with CP2K and a PySCF-based implementation using the aug-cc-pVTZ basis set. For Skala, we additionally considered the atomic-radius adjustment in Becke's integration-grid partitioning, previously identified as a source of differences between CP2K/GauXC and PySCF calculations \cite{Poeschel2026}. PySCF calculations with this adjustment enabled and disabled (denoted as \texttt{radii on} and \texttt{radii off}, respectively) were compared against the same CP2K results. Table~\ref{tab:si_codes} summarizes the mean and maximum absolute differences between the implementations.

\begin{table}[htbp]
\centering
\caption{Mean absolute differences (MADs) and maximum absolute differences (MaxADs) between PySCF and CP2K excitation energies and singlet--triplet gaps at aug-cc-pVTZ (eV). Statistics use 12 values per property and row for QUEST1, 8 for ST-gap, and 11 UKS singlet excitation energies for QUEST6. Radii on/off denotes the atomic-radius adjustment in the Becke integration-grid partitioning used by the PySCF-based Skala implementation; both variants are compared against the same CP2K results. Dashes denote properties outside the reported scope.}
\label{tab:si_codes}
\footnotesize
\setlength{\tabcolsep}{4pt}
\begin{tabular}{@{}llrrrrrr@{}}
\hline
Treatment & XC functional & \multicolumn{2}{c}{$\Delta E_{\mathrm{S}}$} & \multicolumn{2}{c}{$\Delta E_{\mathrm{T}}$} & \multicolumn{2}{c}{$\Delta E_{\mathrm{ST}}$} \\
\cline{3-4}\cline{5-6}\cline{7-8}
 & & MAD & MaxAD & MAD & MaxAD & MAD & MaxAD \\
\hline
\multicolumn{8}{l}{\textit{QUEST1}} \\
UKS & PBE & 0.000 & 0.001 & 0.000 & 0.001 & 0.000 & 0.000 \\
UKS & r$^2$SCAN & 0.002 & 0.004 & 0.003 & 0.007 & 0.003 & 0.008 \\
UKS & PBE0 & 0.000 & 0.001 & 0.000 & 0.001 & 0.000 & 0.000 \\
UKS & Skala (radii on) & 0.013 & 0.035 & 0.008 & 0.030 & 0.009 & 0.035 \\
UKS & Skala (radii off) & 0.002 & 0.004 & 0.001 & 0.003 & 0.001 & 0.004 \\
\noalign{\smallskip}
ROKS & PBE & 0.000 & 0.001 & 0.000 & 0.001 & 0.000 & 0.000 \\
ROKS & r$^2$SCAN & 0.001 & 0.004 & 0.003 & 0.005 & 0.002 & 0.006 \\
ROKS & PBE0 & 0.000 & 0.001 & 0.000 & 0.001 & 0.000 & 0.000 \\
ROKS & Skala (radii on) & 0.013 & 0.051 & 0.008 & 0.031 & 0.008 & 0.026 \\
ROKS & Skala (radii off) & 0.001 & 0.003 & 0.001 & 0.002 & 0.001 & 0.002 \\
\hline
\multicolumn{8}{l}{\textit{ST-gap}} \\
UKS & PBE & 0.001 & 0.001 & 0.001 & 0.001 & 0.000 & 0.000 \\
UKS & r$^2$SCAN & 0.009 & 0.010 & 0.004 & 0.005 & 0.005 & 0.006 \\
UKS & PBE0 & 0.001 & 0.001 & 0.001 & 0.001 & 0.000 & 0.001 \\
UKS & Skala (radii on) & 0.006 & 0.012 & 0.006 & 0.010 & 0.002 & 0.004 \\
UKS & Skala (radii off) & 0.001 & 0.002 & 0.001 & 0.003 & 0.001 & 0.001 \\
\noalign{\smallskip}
ROKS & PBE & 0.001 & 0.001 & 0.001 & 0.001 & 0.000 & 0.000 \\
ROKS & r$^2$SCAN & 0.003 & 0.004 & 0.003 & 0.004 & 0.000 & 0.000 \\
ROKS & PBE0 & 0.001 & 0.001 & 0.001 & 0.001 & 0.000 & 0.000 \\
ROKS & Skala (radii on) & 0.006 & 0.015 & 0.006 & 0.010 & 0.003 & 0.004 \\
ROKS & Skala (radii off) & 0.001 & 0.003 & 0.001 & 0.003 & 0.001 & 0.001 \\
\hline
\multicolumn{8}{l}{\textit{QUEST6}} \\
UKS & PBE & 0.004 & 0.010 & -- & -- & -- & -- \\
UKS & r$^2$SCAN & 0.005 & 0.018 & -- & -- & -- & -- \\
UKS & PBE0 & 0.005 & 0.018 & -- & -- & -- & -- \\
UKS & Skala (radii on) & 0.005 & 0.024 & -- & -- & -- & -- \\
UKS & Skala (radii off) & 0.005 & 0.014 & -- & -- & -- & -- \\
\hline\hline
\end{tabular}
\end{table}

For QUEST1 and ST-gap, PBE and PBE0 show close agreement between the implementations, with maximum deviations not exceeding 0.001~eV. Somewhat larger differences are observed for r$^2$SCAN, reaching 0.010~eV for ST-gap. For Skala, the maximum deviations for QUEST1 and ST-gap are 0.051 and 0.015~eV, respectively, with \texttt{radii on}, decreasing to 0.004 and 0.003~eV with \texttt{radii off}. Notably, this close agreement between CP2K and PySCF-based calculations is achieved despite differences in the SCF optimization procedures and, for some systems, the initial orbitals, providing additional evidence of the numerical robustness of the calculated excitation energies.

For QUEST1 and ST-gap, PBE and PBE0 show close agreement between the implementations, with maximum deviations not exceeding 0.001~eV. Somewhat larger differences are observed for r$^2$SCAN, reaching 0.010~eV for ST-gap. For Skala, the maximum deviations for QUEST1 and ST-gap are 0.051 and 0.015~eV, respectively, with \texttt{radii on}, decreasing to 0.004 and 0.003~eV with \texttt{radii off}. Notably, this close agreement is obtained despite differences in the SCF optimization procedures and the initialization strategies.

\clearpage
\section{Functional substitution and orbital relaxation in ROKS}
\label{sec:si_roks_frozen_orbitals}

To examine the functional and orbital contributions to the ST-gap results, we evaluated PBE and Skala on orbital sets optimized with each functional. All calculations use CP2K with aug-cc-pVTZ at fixed benchmark geometries. We denote an evaluation with functional $F$ on orbitals optimized with $G$ by $F@G$. The singlet and restricted-triplet orbital sets, $C_{\mathrm S}^{G}$ and $C_{\mathrm T}^{G}$, were optimized separately with $G$ and held fixed during evaluation with $F$. Using the ROKS singlet-energy expression, Eq.~\eqref{eq:roks_singlet_energy}, the corresponding gap is
\begin{equation}
\Delta E_{\mathrm{ST}}(F@G)
=2E_{\mathrm M}^{F}(C_{\mathrm S}^{G})
-E_{\mathrm T}^{F}(C_{\mathrm S}^{G})
-E_{\mathrm T}^{F}(C_{\mathrm T}^{G}).
\label{eq:si_frozen_gap}
\end{equation}
Thus, each cross-functional gap uses the evaluating functional for all three component energies, including the triplet energy on the separately optimized donor orbitals. No further orbital optimization is performed in a cross-evaluation.

\begin{table}[htbp]
\centering
\caption{Frozen-orbital ROKS singlet--triplet gaps for ST-gap (CP2K, aug-cc-pVTZ; eV). $F@G$ denotes evaluation with $F$ on the respective singlet and separately optimized restricted-triplet orbitals obtained with $G$. The four columns contain 0, 3, 0, 6 negative gaps, respectively, among the eight molecules. Both functional/orbital comparison paths are retained; their contributions are path-dependent.}
\label{tab:si_roks_frozen_orbitals}
\footnotesize
\setlength{\tabcolsep}{5pt}
\begin{tabular}{@{}lrrrr@{}}
\hline\hline
Molecule & PBE@PBE & Skala@PBE & PBE@Skala & Skala@Skala \\
\hline
Heptazine & +0.109 & +0.001 & +0.135 & \textminus{}0.025 \\
Cyclazine & +0.109 & \textminus{}0.033 & +0.149 & \textminus{}0.071 \\
Molecule~3 & +0.125 & \textminus{}0.022 & +0.164 & \textminus{}0.062 \\
Molecule~4 & +0.131 & +0.005 & +0.159 & \textminus{}0.022 \\
Molecule~5 & +0.132 & \textminus{}0.009 & +0.170 & \textminus{}0.047 \\
Molecule~6 & +0.133 & +0.022 & +0.169 & \textminus{}0.009 \\
Molecule~7 & +0.147 & +0.066 & +0.177 & +0.042 \\
Molecule~8 & +0.156 & +0.097 & +0.180 & +0.079 \\
\hline
Mean & +0.130 & +0.016 & +0.163 & \textminus{}0.014 \\
\hline\hline
\end{tabular}
\end{table}

Table~\ref{tab:si_roks_frozen_orbitals} gives all four functional/orbital combinations. Along the path PBE@PBE $\to$ Skala@PBE $\to$ Skala@Skala used in Fig.~\ref{fig:st_gap_roks_analysis}, functional substitution lowers the mean gap by 0.114~eV, followed by a further decrease of 0.030~eV upon orbital optimization with Skala. Cyclazine and Molecules~3 and~5 already have negative gaps at Skala@PBE; Heptazine and Molecules~4 and~6 become inverted after orbital optimization. 

The alternative path, PBE@PBE $\to$ PBE@Skala $\to$ Skala@Skala, instead raises the mean gap by 0.033~eV when the orbital sets are changed at fixed PBE and then lowers it by 0.177~eV when PBE is replaced by Skala at fixed Skala orbitals. Both paths connect the same endpoints and give a total mean reduction of 0.145~eV, but assign different changes to the individual steps. These quantities therefore describe the chosen comparison path rather than a unique decomposition of functional and orbital effects.

\clearpage
\section{Effect of separate triplet optimization on ROKS gaps}
\label{sec:si_triplet_relaxation}

To determine whether separate triplet optimization is required for the ROKS inversions, we compare common-orbital splittings and gaps between separately optimized states for PBE, r$^2$SCAN, PBE0, and Skala. For a given functional, let $C_{\mathrm S}$ and $C_{\mathrm T}$ denote the orbital sets optimized for the ROKS singlet and restricted open-shell triplet, respectively. All energies below use that same functional at a fixed nuclear geometry. In Eq.~\eqref{eq:roks_singlet_energy}, the mixed-spin and triplet components are both evaluated on $C_{\mathrm S}$. Using this triplet component without separate triplet optimization gives the common-orbital splitting
\begin{equation}
\Delta E_{\mathrm{ST}}^{\mathrm{common}}
=E_{\mathrm S}(C_{\mathrm S})-E_{\mathrm T}(C_{\mathrm S})
=2\bigl[E_{\mathrm M}(C_{\mathrm S})-E_{\mathrm T}(C_{\mathrm S})\bigr].
\label{eq:si_common_gap}
\end{equation}
The singlet orbitals are fully optimized in this evaluation; only the separate optimization of the triplet is omitted. The resulting splitting is therefore distinct from a gap between independently optimized states. Defining the triplet-relaxation correction as
\begin{equation}
\delta_{\mathrm T}=E_{\mathrm T}(C_{\mathrm S})-E_{\mathrm T}(C_{\mathrm T}),
\label{eq:si_triplet_correction}
\end{equation}
the gap obtained with the separately optimized triplet is
\begin{equation}
\Delta E_{\mathrm{ST}}^{\mathrm{opt}}
=E_{\mathrm S}(C_{\mathrm S})-E_{\mathrm T}(C_{\mathrm T})
=\Delta E_{\mathrm{ST}}^{\mathrm{common}}+\delta_{\mathrm T}.
\label{eq:si_relaxed_gap}
\end{equation}
The singlet energy is identical in the two gap evaluations, and no ground-state energy is required. On the common orbital set, the mixed and triplet determinants have the same total density, and the kinetic, external-potential, and Hartree contributions cancel. We verified this cancellation numerically, including the GAPW contributions. The common-orbital splitting is consequently determined by twice the difference between their XC energies, including exact exchange for PBE0.

\begin{table}[htbp]
\centering
\caption{Effect of separate triplet optimization on ROKS singlet--triplet gaps for ST-gap (CP2K, aug-cc-pVTZ; eV). Common: $E_{\mathrm S}(C_{\mathrm S})-E_{\mathrm T}(C_{\mathrm S})$; optimized: $E_{\mathrm S}(C_{\mathrm S})-E_{\mathrm T}(C_{\mathrm T})$, where $C_{\mathrm S}$ and $C_{\mathrm T}$ are singlet- and restricted-triplet-optimized orbitals. The singlet and nuclear geometry remain fixed. Skala gives 7 common-orbital and 6 optimized-gap inversions among the eight molecules. No inversions occur for the other functionals.}
\label{tab:si_triplet_relaxation}
\footnotesize
\setlength{\tabcolsep}{4pt}
\begin{tabular}{@{}lrrrrrrrr@{}}
\hline\hline
Molecule & \multicolumn{2}{c}{PBE} & \multicolumn{2}{c}{r$^2$SCAN} & \multicolumn{2}{c}{PBE0} & \multicolumn{2}{c}{Skala} \\
\cline{2-3}\cline{4-5}\cline{6-7}\cline{8-9}
 & Common & Optimized & Common & Optimized & Common & Optimized & Common & Optimized \\
\hline
Heptazine & +0.101 & +0.109 & +0.088 & +0.103 & +0.124 & +0.139 & \textminus{}0.053 & \textminus{}0.025 \\
Cyclazine & +0.100 & +0.109 & +0.089 & +0.105 & +0.122 & +0.134 & \textminus{}0.126 & \textminus{}0.071 \\
Molecule~3 & +0.114 & +0.125 & +0.109 & +0.130 & +0.147 & +0.177 & \textminus{}0.115 & \textminus{}0.062 \\
Molecule~4 & +0.119 & +0.131 & +0.115 & +0.141 & +0.152 & +0.191 & \textminus{}0.064 & \textminus{}0.022 \\
Molecule~5 & +0.119 & +0.132 & +0.115 & +0.142 & +0.158 & +0.204 & \textminus{}0.099 & \textminus{}0.047 \\
Molecule~6 & +0.120 & +0.133 & +0.118 & +0.145 & +0.149 & +0.176 & \textminus{}0.071 & \textminus{}0.009 \\
Molecule~7 & +0.134 & +0.147 & +0.139 & +0.168 & +0.165 & +0.193 & \textminus{}0.021 & +0.042 \\
Molecule~8 & +0.142 & +0.156 & +0.154 & +0.181 & +0.175 & +0.196 & +0.017 & +0.079 \\
\hline
Mean & +0.119 & +0.130 & +0.116 & +0.139 & +0.149 & +0.176 & \textminus{}0.066 & \textminus{}0.014 \\
\hline\hline
\end{tabular}
\end{table}

Table~\ref{tab:si_triplet_relaxation} shows positive common-orbital splittings for all eight molecules with PBE, r$^2$SCAN, and PBE0, with means of 0.119, 0.116, and 0.149~eV, respectively. Separate triplet optimization increases the gaps for every molecule, giving mean gaps of 0.130, 0.139, and 0.176~eV. Skala instead gives negative common-orbital splittings for seven molecules, with a mean of \textminus{}0.066~eV. Separate triplet optimization increases its gaps by 0.052~eV on average; six remain negative, with a mean gap of \textminus{}0.014~eV. Only Molecule~7 changes sign, from \textminus{}0.021 to +0.042~eV. The Skala inversions are therefore already present on the singlet-optimized orbital sets and do not require separate triplet relaxation. This comparison holds each functional and its singlet state fixed, whereas the orbital-change steps in Sec.~\ref{sec:si_roks_frozen_orbitals} change both singlet and triplet orbital sets.